\documentstyle[12pt,blois, psfig]{article}

\begin{document}

\heading{METALLIC ABSORBERS AND THE EVOLUTION OF GALAXY HALOES AND THE 
METAGALACTIC IONIZING BACKGROUND}

\author{Alec Boksenberg $^{1}$, Wallace L.W. Sargent $^{2}$ \& Michael Rauch $^{2}$} {$^{1}$ University of Cambridge, Institute of Astronomy, 
Madingley Road, Cambridge, UK.}  {$^{2}$ Astronomy Dept., 105-25 California Institute of Technology, Pasadena, CA 91125, USA.}

\begin{bloisabstract}
An extremely careful separation of the weak metal-line systems
relating to the Lyman forest absorbers into individual ``single-phase
ionization'' components with accurate parameters has yielded a large
sample in the wide redshift range $1.9 < z < 4.4$.  The systems
typically span several hundred km s$^{-1}$ and within each the
components show a strong coherence in their ionization properties.  No
sudden evolution is found in the column density ratio N(Si IV)/N(C
IV), contrary to previous indication of a large change near $z = 3$,
although there is an apparent rise below $z \sim 2.2$ and above $z
\sim 3.8$.  A smooth not sudden spectral evolution is reflected in the
$z$ distribution of N(N V)/N(C IV).  Comparisons of N(Si IV)/N(C IV)
vs N(C II)/N(C IV) from the observations with CLOUDY derived values
using several model ionizing spectra show the general dominance of a
QSO ionizing background at the lower redshifts while an additional
strong and eventually dominating stellar contribution is required
progressively to higher redshifts.
\end{bloisabstract}

\section{Introduction}
Until quite recently most astronomers believed the Lyman forest clouds
to be pristine.  It was a surprise, therefore, when the spectra of
exceptional quality delivered by the high resolution spectrograph of
the Keck Telescopes revealed individual metal features related to the
high redshift Lyman forest for a large fraction of the stronger lines
\cite{20, 21, 22}.  These metal systems provide a powerful probe of
the earliest stages in the growth of structure and the formation of
galaxies and give an observational approach to determining the
character of the cosmological ionizing sources and the stellar
birthrate at those times.  Here we give some new results on such metal
systems from an analysis of the high quality Keck spectra of the 9
QSOs (emission redshift in brackets) 1626 + 6433 (2.32), 1442 + 2931
(2.67), 1107 + 4847 (2.97), 0636 + 6801 (3.18), 1425 + 6039 (3.20), 1422
+ 2309 (3.62), 1645 + 5520 (4.06), 1055 + 4611 (4.13) and 2237 - 0607
(4.56), obtained at a resolution $\sim 6.3$ km s$^{-1}$ FWHM.  This
work will be more fully described and discussed in a forthcoming
publication (in preparation).

\section{Component Structure}

In general appearance the metal systems associated with the Lyman
forest lines are blended complexes of absorber components.  They
display a range of ionized species, where detectable in the overall
spectrum, from the weakest in which only C IV is seen to the strongest
which also contain Si IV, C II, Si II, Al II, ${\rm Al\ III}$, O I, Fe
II, Ni II and frequently N V.  Typically a complex spans a few hundred
km s$^{-1}$ and often is associated with one or more others in a close
group with wide expanses of clear spectrum between such groups.  In an
extremely careful analysis using the Voigt profile fitting package
VPFIT \cite{23}, such complexes, involving all detected ionic species,
here have been self-consistently separated into individual component
``clouds'' closely approximating single-phase ionization regions with
quite accurate values for column density (N), Doppler parameter (b)
and redshift ($z$).  Excellent simultaneous fits over all species are
achieved with the same pattern of redshifts in each and the same
pattern of b-values for all ions of a given atom.  Most of the
components are narrow with b(C IV) largely contained within the range
4-10 km s$^{-1}$.  Particularly for the stronger components (which are
more precisely determined) the resultant ratios of b-values for ions
of different atoms, e.g., ${\rm b(Si\ IV)/b(C\ IV)}$, are physically realistic
and yield a temperature structure within the complexes containing
values distributed up to a few $\times 10^4$K, typical of photoionization
heating.

A frequent feature of the systems is the presence of some broad (a few
$\times 10$ km s$^{-1}$), high ionization components which co-exist in
velocity space with the much more numerous narrow components of lower
ionization \cite{29}.  Such sets of different component structures
must therefore be signatures of physically distinct but spatially
closely related regions.  While the implicit model in such profile
constructions has only temperature and Gaussian turbulence broadening
included in the specific b-parameter characterising each assumed cloud
in a complex, large velocity gradients from bulk motions must also
contribute to the true overall absorption profile.  The broad, high
ionization components revealed thus probably represent regions of low
volume density dominated by bulk motions.

\begin{figure}[t]
\vskip -.74cm \hbox{ \hskip 0cm{
\vbox{\psfig{figure=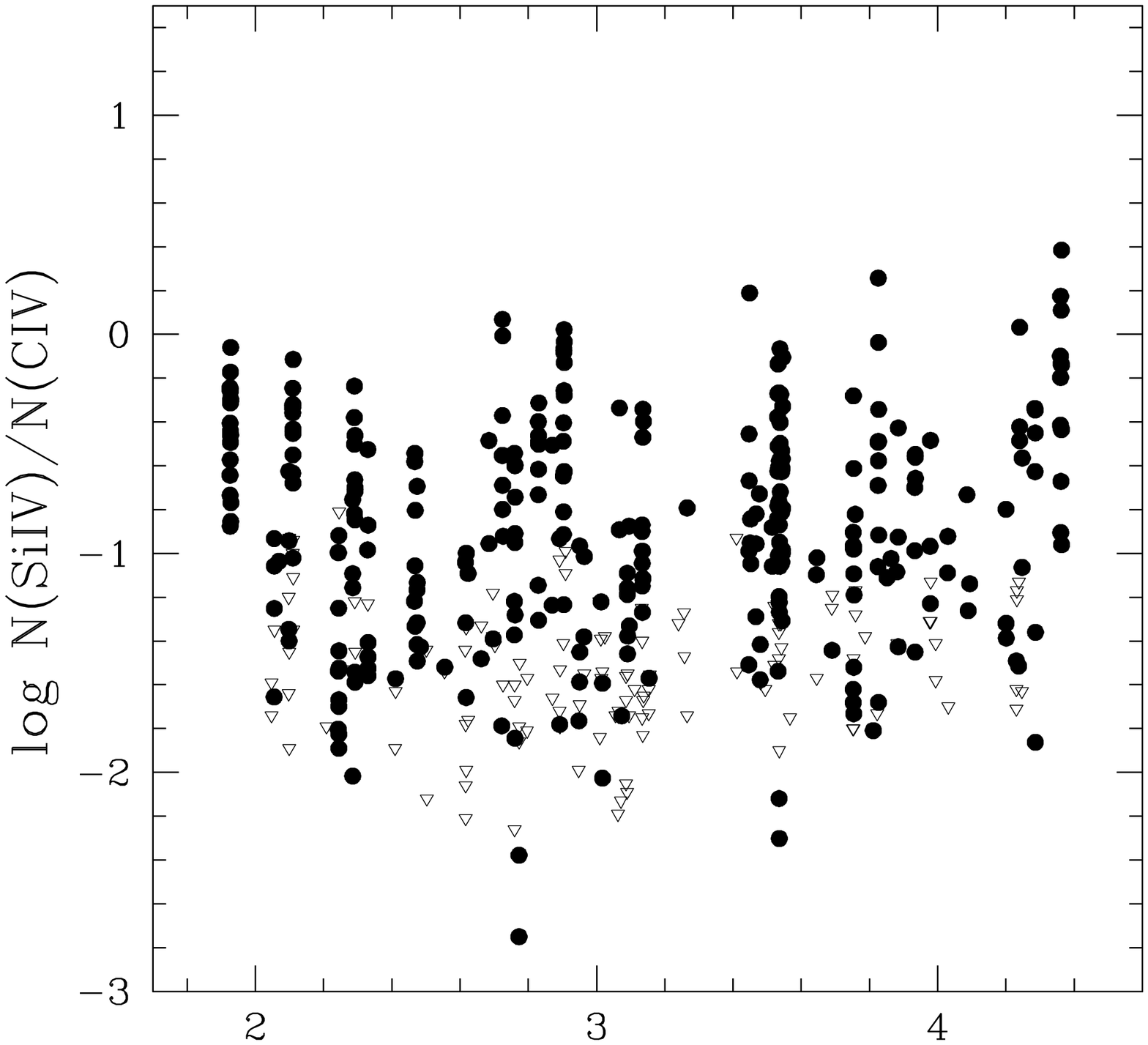,width=8.4cm}\vskip -.9cm
\psfig{figure=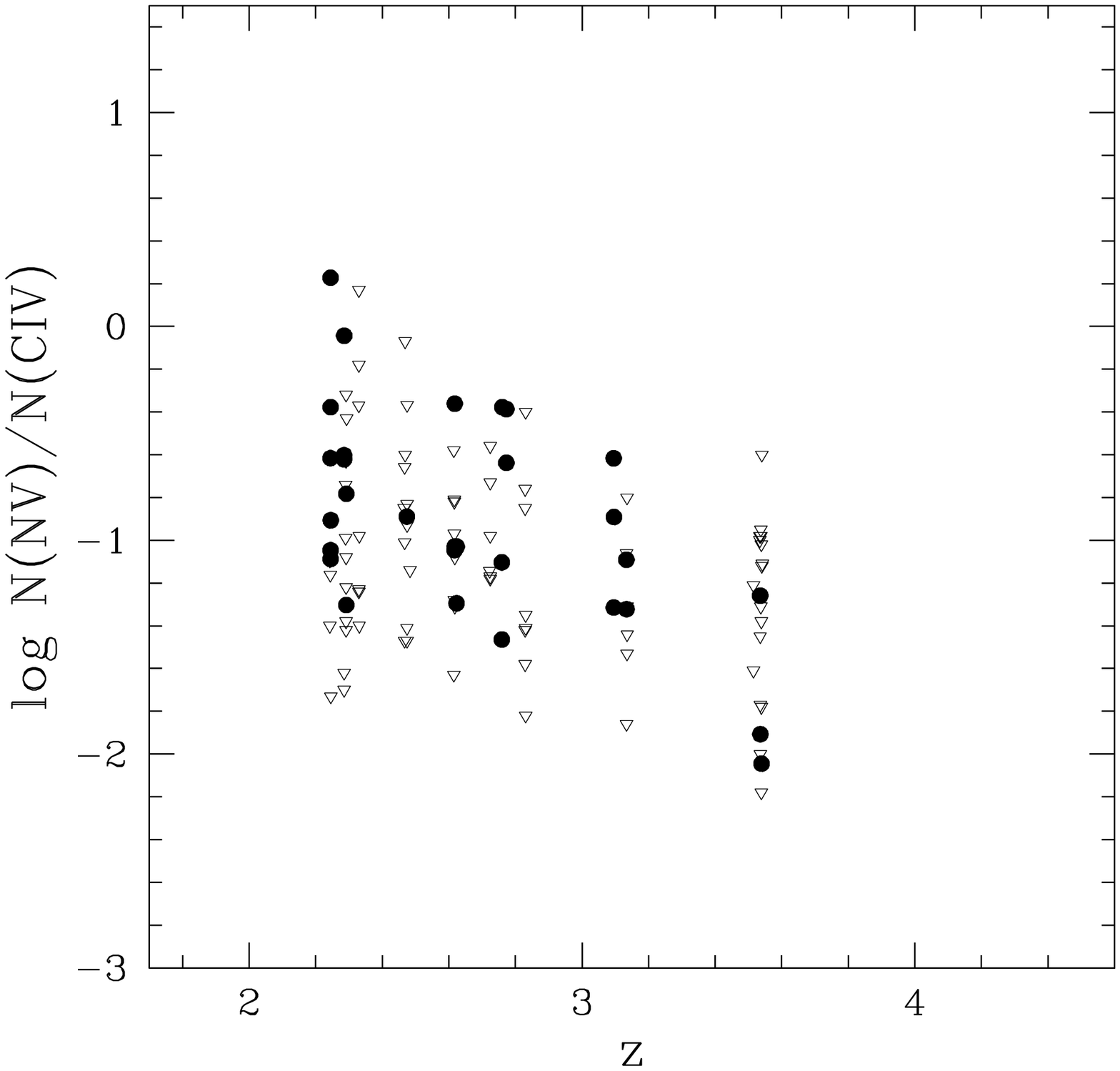,width=8.4cm}}} 
\vbox{\psfig{figure=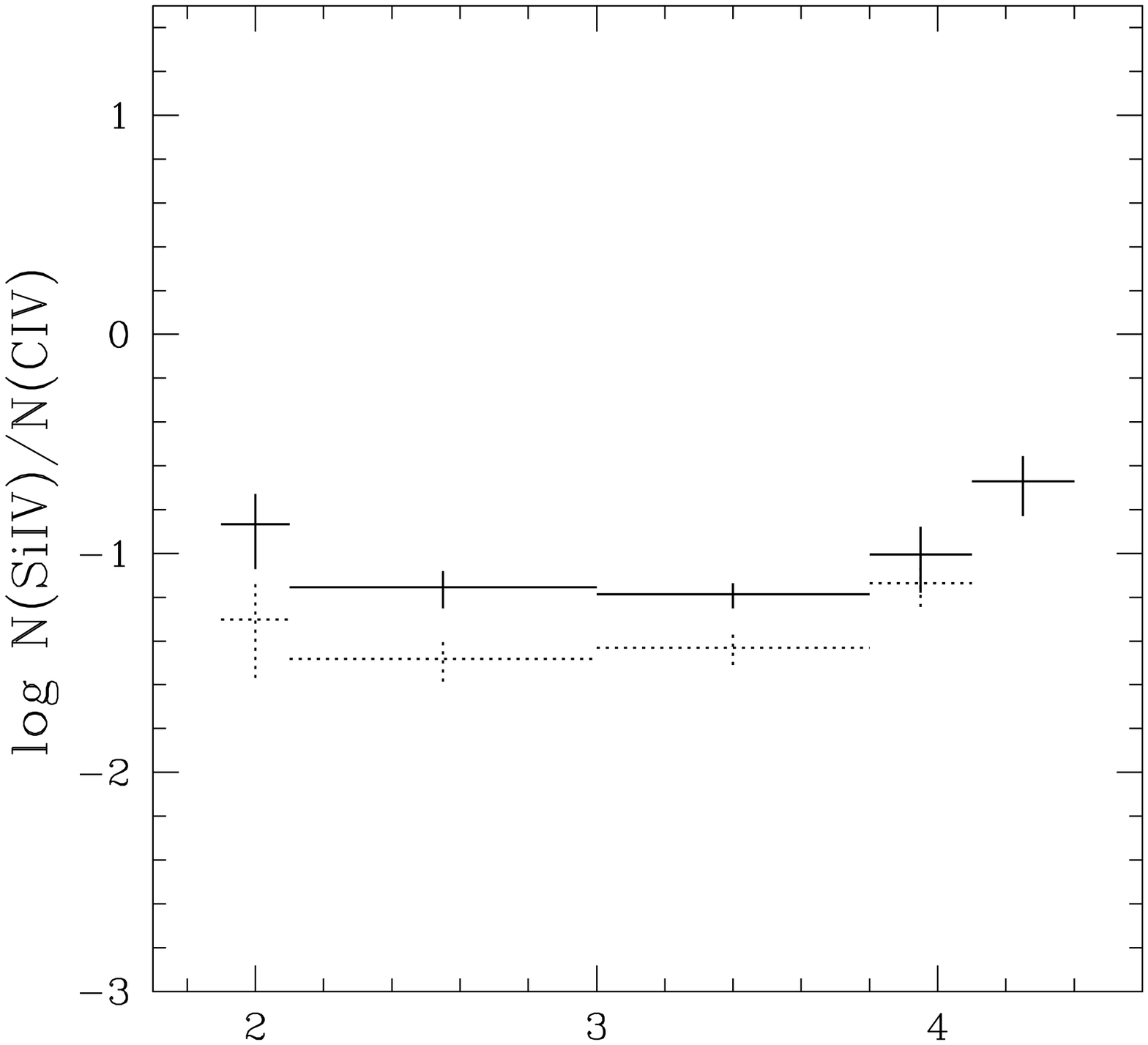,width=8.4cm}\vskip -.9cm
\psfig{figure=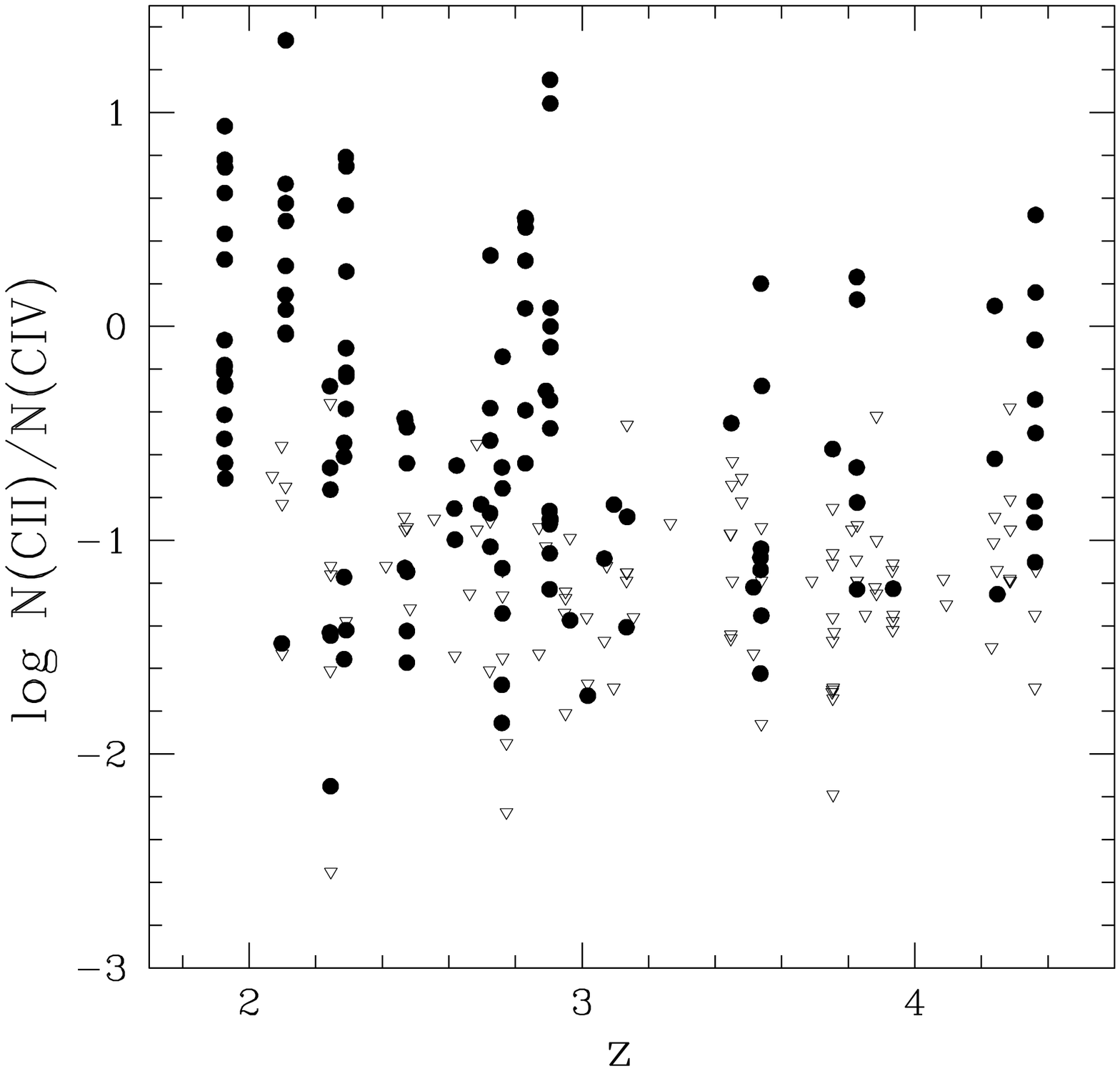,width=8.4cm}}}\vskip -0.2cm
\caption{{\small{{\it Upper left} -- N(Si IV)/N(C IV) vs redshift
($z$) for all cloud components in the spectra of 9 QSOs with Si IV
outside the Lyman forest and N(C IV) $\geq 10^{12}$ cm$^{-2}$.
Inverted open triangles are values with 1-$\sigma$ upper limits
for N(Si IV).  All systems within 3000 km s$^{-1}$ of the lowest QSO
redshifts and progressively up to 10000 kms$^{-1}$ of the highest
redshifts, and systems assessed as not optically thin in the Lyman
continuum, are excluded.  The components of each absorption complex,
separated as described in the text, show here as vertically
distributed associations (several complexes, from different QSOs,
happen to overlap near $z = 3.5$).  {\it Upper right} -- Median values
for the data at {\it left} describing the full data set (full line)
and only the simple systems (5 or fewer components -- dotted line).
Errors are $\pm 1$-$\sigma$ values. {\it Lower left} -- N(N V)/N(C IV)
as for N(Si IV)/N(C IV) with the exception that many N(V) values are
obtained from components within the Lyman forest where it happens to
be relatively clear; apart from strong, unambiguous cases these values
are taken as upper limits. {\it Lower right} -- N(C II)/N(C IV) as for
N(Si IV)/N(C IV).  }}}
\end{figure}

\section{Ionization Balance and the Ionizing Radiation Environment}

Values of the column density ratio N(Si IV)/N(C IV) among the
components in each system show remarkable coherence, typically
extending over a factor only $\sim 10$.  There are similar bulk
differences between systems.  It is interesting that such an
ionization pattern is not predicted in recent hydrodynamical
simulations of metal-bearing collapsing gas structures in the Universe
photoionized by a metagalactic ionizing background \cite{12}, although
in many other aspects this modeling shows considerable success in
reproducing the observed characteristics of absorbers.  

As shown in Figure 1 the distribution in redshift of ${\rm N(Si\
IV)/N(C\ IV)}$ for the components of all systems detected over the
range $z = 1.9 - 4.4$ (but sufficiently removed from the target QSOs
to avoid proximity effects and excluding those assessed as not
optically thin in the Lyman continuum to avoid complications from
self-shielding) is smooth in the observed range, with a
progressive rise in the median value up to a factor $\sim 3$ above $z
\sim 3.8$ and $\sim 2$ below $z \sim 2.2$, and is quite constant
between these redshifts.  Sub-sets of the data, including those
containing only simple systems (5 or fewer components) and only
complex systems (6 or more components), all show similar trends.
Median values for the simple systems also are shown in Figure 1; that
these are somewhat lower than those of the full data set (and of the
complex systems) is consistent with the positions of these generally
more highly ionized systems on the diagrams to be described in Section
3.  This observed trend with redshift is quite contrary to previous
presentations of a sudden large change near $z =3$, interpreted as due
to a change in the metagalactic ionizing spectrum at that epoch from a
sudden reduction in the opacity of the evolving intergalactic medium
to He$^+$ ionizing photons as He$^+$ ionizes completely to He$^{++}$
\cite{24,25}.  Although the basis of the difference is not clear, it
is important to stress that the Voigt profile component-fitting
technique used here has much more physical relevance than using
aggrated system quantities \cite{24}, which mix the ionization
conditions, or optical depth determinations \cite{25}, which do not
discriminate against blends, overlapping structures or temperature
effects.  While the relevant ionization potentials of the Si and C
species related to the appearance of Si IV and C IV straddle the
He$^+$ edge, the presence of N(V) requires higher energies so is a
more sensitive indicator of He$^+$ continuum photons.  The redshift
distribution of N(N V)/N(C IV) (Figure 1) shows smooth not sudden
spectral evolution rising by a factor $\sim 10$ continuously from $z \sim
3.5$ to $z \sim 2.0$.  This is consistent with a continuously diminishing
break at the He$^+$ edge, not only from progressive evolution in the
opacity of the intergalactic medium but, importantly, also from a
greater effective presence of stellar systems, with their
intrinsically soft emission spectra, over QSOs at higher redshifts.
Also shown in Figure 1 is the distribution of ${\rm N(C\ II)/N(C\
IV)}$, closely representing the ionization parameter.  This also has a
general rising trend with decreasing redshift, here by a factor $\sim
4$ from $z \sim 4$, with an apparent rise also to higher redshift
beyond $z \sim 4$.

\begin{figure}[t]
\vskip -.2cm \hbox{ \hskip 0cm{
\vbox{\psfig{figure=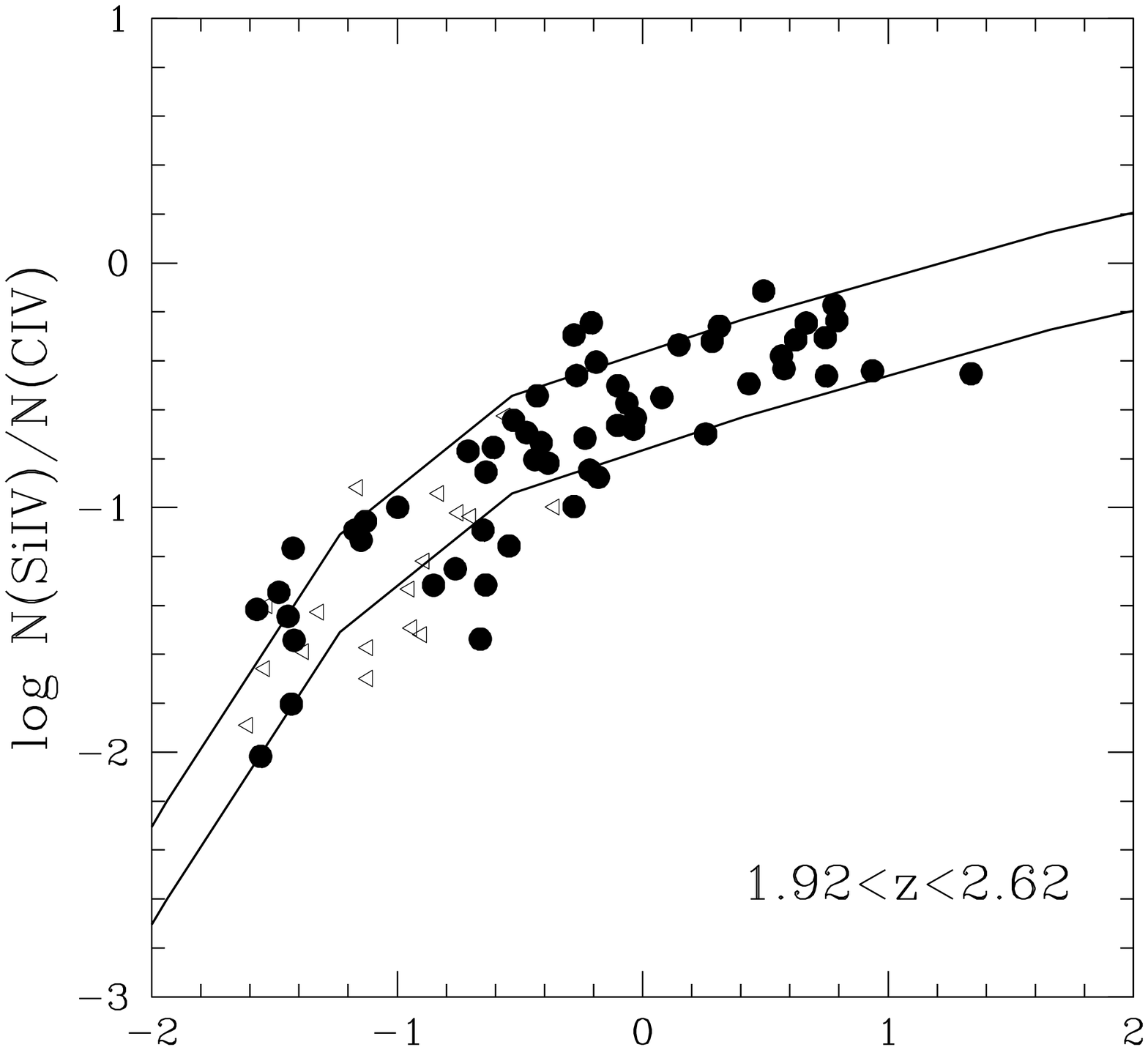,width=5.8cm}\vskip -.4cm
\psfig{figure=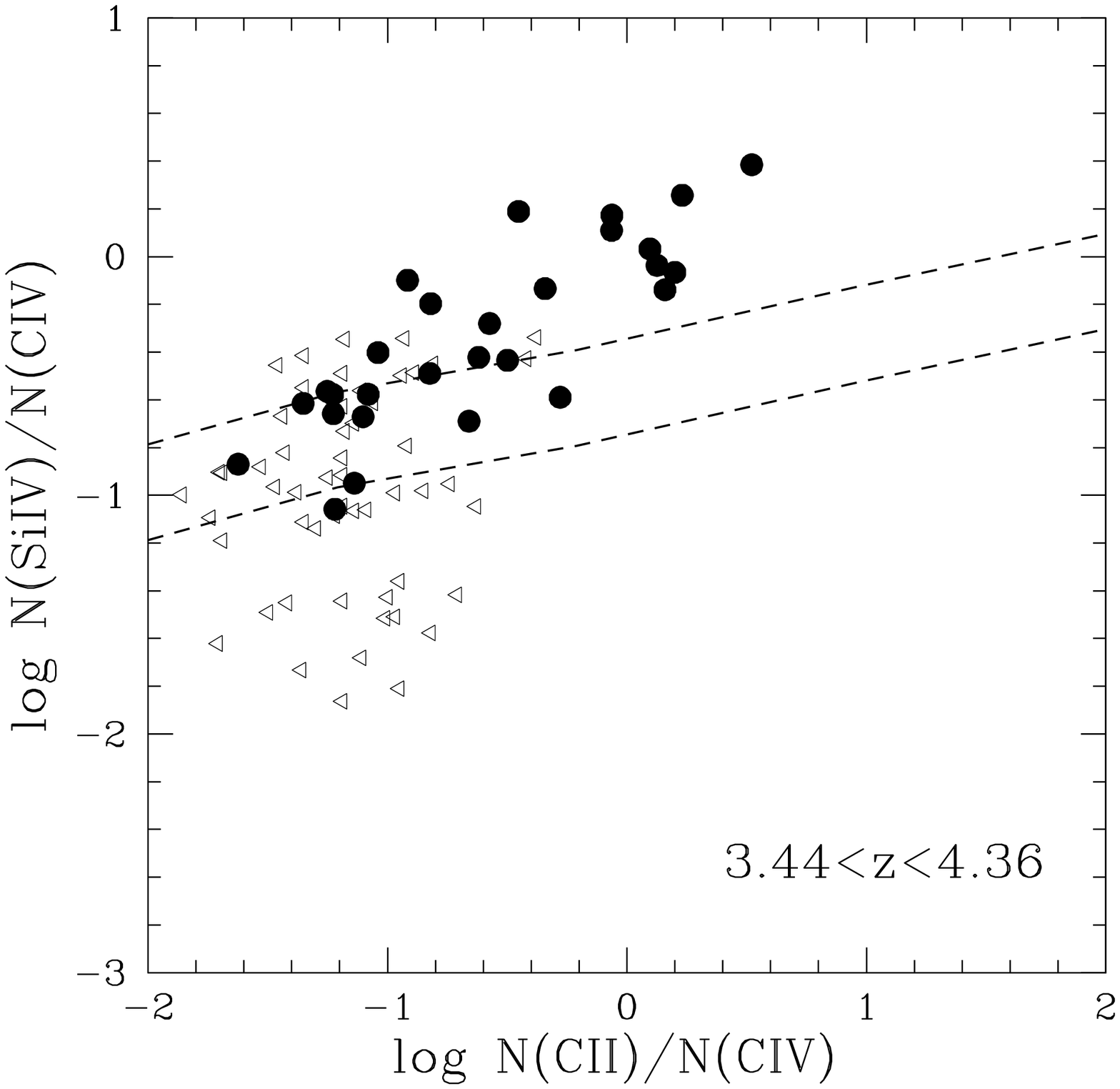,width=5.8cm}}} \hskip -.5cm 
\vbox{\psfig{figure=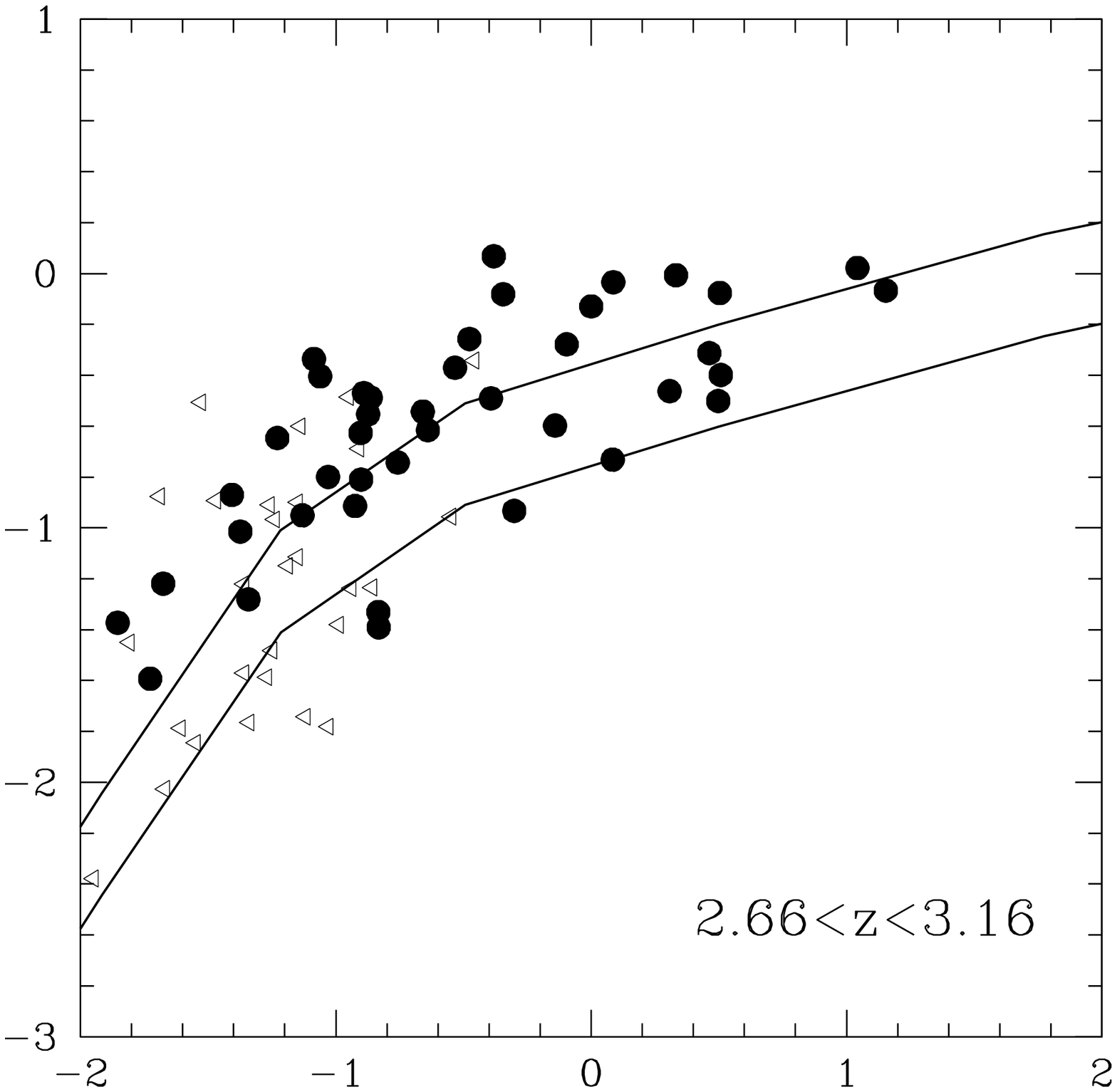,width=5.8cm}\vskip -.4cm
\psfig{figure=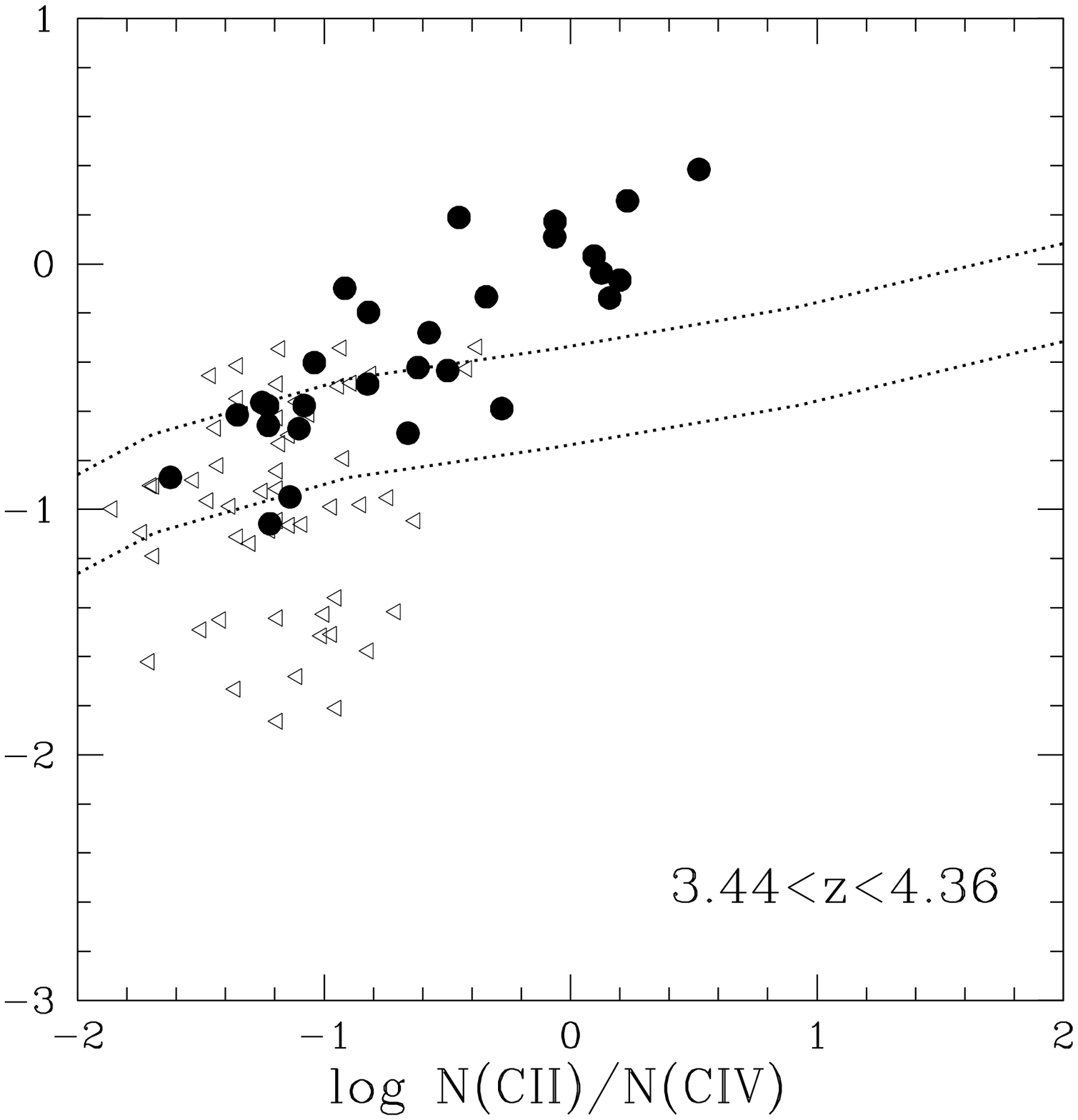,width=5.8cm}} \hskip -.5cm
\vbox{\psfig{figure=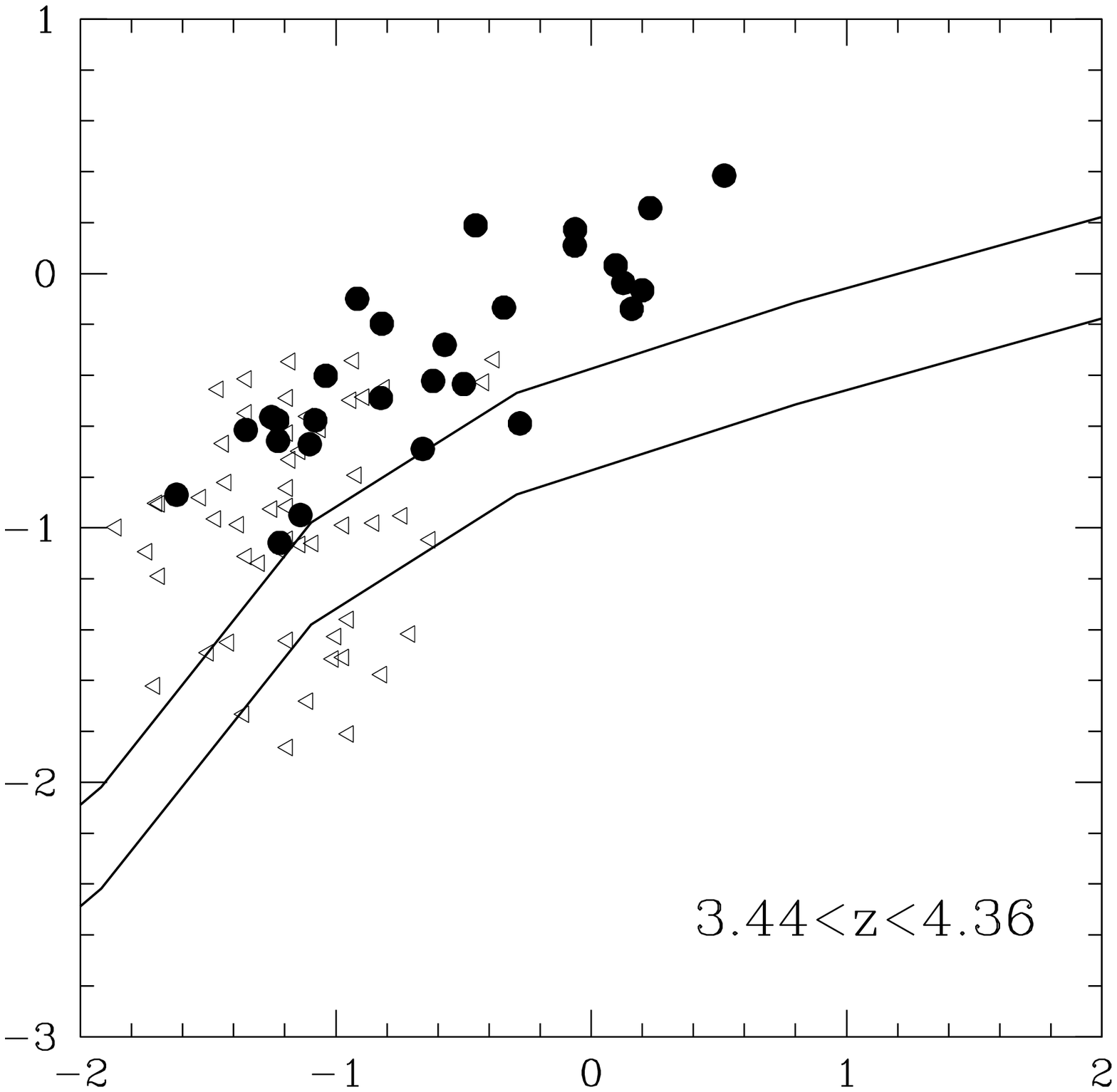,width=5.8cm}\vskip -.4cm
\psfig{figure=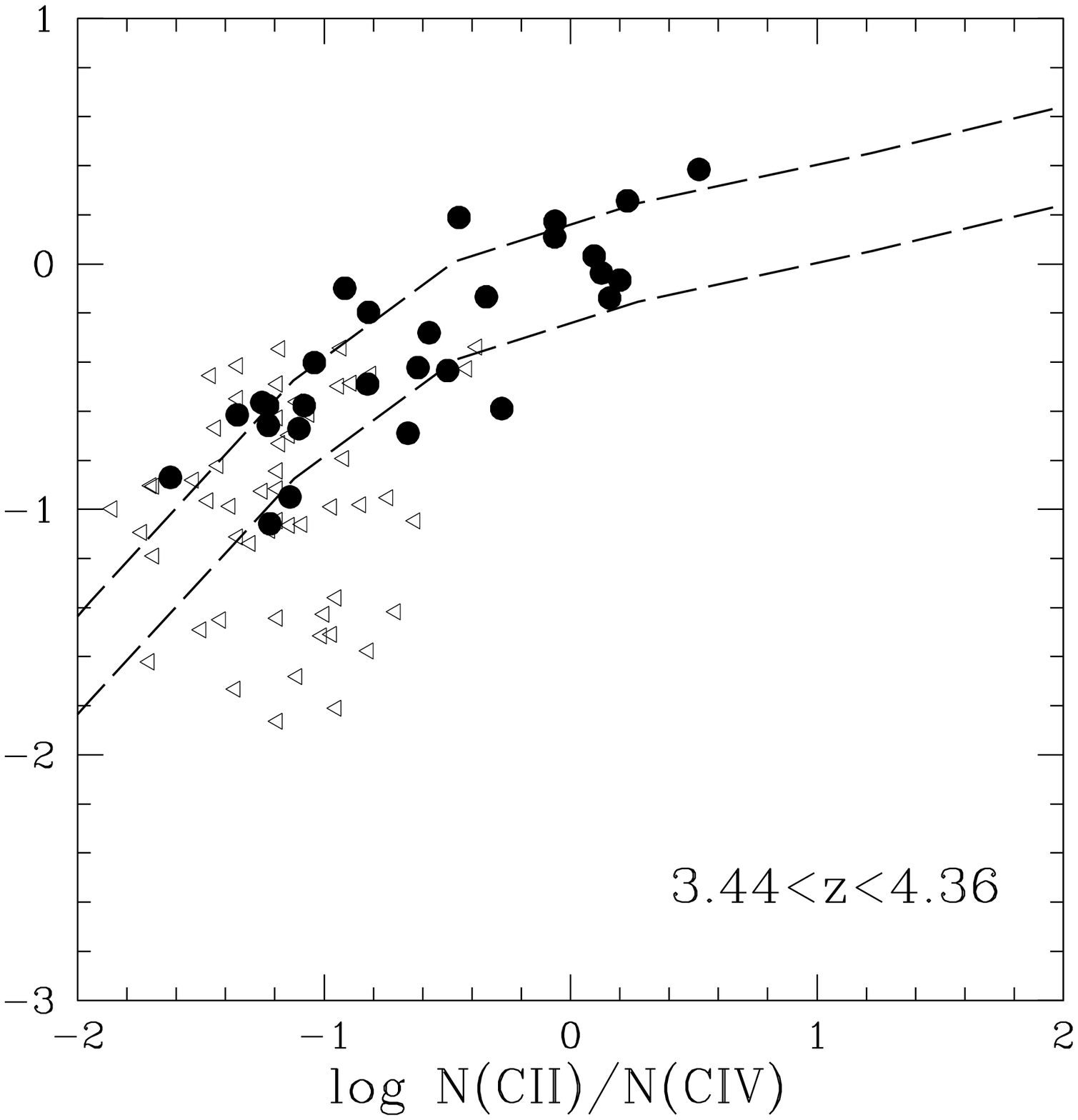,width=5.8cm}}}
\caption{{\small{Comparison of cloud component ion column density
ratios with C II and Si IV outside the Lyman forest and N(CIV) $\geq
10^{12}$ cm$^{-2}$ from the spectra of 9 QSOs, with model predictions
of the CLOUDY code in the optically thin regime and for low
metallicity (1/300 $\times$ solar), for three redshift ranges.
Left-indicating open triangles are values with 1-$\sigma$ upper limits
for C II.  Upper limits including both C II and Si IV (not shown) all are
consistent with the data.  {\it Upper panels} -- Model predictions
computed for Haardt \& Madau \cite{27} latest available versions for the
QSO UV background \cite{28} appropriate for the indicated
redshifts using $q_0 = 0.5$ and in each showing the two cases of Si/C
relative abundance: solar (lower line) and $2.5 \times$ solar (upper
line).  {\it Lower panels} -- Model predictions computed for the
highest redshift range ({\it upper right}) for three different contributions to the
ionizing flux at the absorbers, again showing the two cases of Si/C
relative abundance: {\it short-dash lines} -- the same QSO background
as in the upper panel but with a cut-off beyond 4 Rydbergs; {\it
dotted lines} -- the full QSO background with an additional
contribution from a source population of star-forming galaxies (see text)
included in the radiative transfer analysis, equal to 30 times the
general QSO flux at 1 Rydberg \cite{50}; {\it long-dash line} --
the full QSO background with an additional local stellar contribution
(Kurucz 45000K, ${\rm log}\ g = 4.5$ \cite{35}, available within the
CLOUDY code) equal to 30 times the general QSO flux at 1 Rydberg.  The
cosmic microwave background, a significant cause of Compton cooling
for low density clouds, is included in all cases.}}}
\end{figure}

Displays of ${\rm N(Si\ IV)/N(C\ IV)}$ vs N(C II)/N(C IV) give much
more complete information on the spectral shape of the ionizing flux.
Figure 2 compares values for these ratios, derived for the individual
components in the full data set, with model predictions of the CLOUDY
code \cite{26} (version 90.04) computed for several trial ionizing
fields with the aim of matching the observed characteristic curves
traced by the observed ion ratios.  For a lower redshift set of the
data $(z = 1.9 - 2.6)$ a good match is achieved with (a) an ionizing
spectrum contributed by QSOs and reprocessed by the intergalactic
medium \cite{27, 28} and (b) low metallicity clouds with a
distribution in the relative abundance of silicon to carbon ranging
over $\sim 1 - 2.5$ times the solar value, similar to that exhibited
by stars in the Galactic halo.  Towards higher redshifts there is
an increasing overall rise in the data, by a factor $\sim 2$ for a set
at $z = 2.7 - 3.2$ and $\sim 4$ for a set at $z = 3.4 - 4.4$.  As
already noted there is a general accompanying trend in the
distribution in ${\rm N(C\ II)/N(C\ IV)}$ towards lower values (higher
ionization parameter), implying greater intensity of the ionizing flux
with increasing redshift, lower absorber density, or both.  As shown
by the CLOUDY modelling in Figure 2 this observed change with
increasing redshift cannot be explained by increasing (albeit not
sudden) opacity of the intergalactic medium at the He$^+$ ionization
edge.  Also tried was a metagalactic ionizing
background computed with increasing source contributions from
star-forming galaxies adding to the QSO contribution \cite{50} (with
the QSOs following the observed drop-off with redshift \cite{27}).  A
specific case is shown using a model selected from the 1996 updated
version of the Bruzual \& Charlot population synthesis code
\cite{51,52} (Salpeter IMF, constant star-formation rate, age $3
\times 10^8$ years).  In both these trials there is a substantial rise
in the computed characteristic curves only at higher ionization
levels, not a general rise as in the observations.  Achieving a
general rise is very sensitive to the spectral shape between $\sim
2.5$ and 4 Rydbergs.  Contributions to the metagalactic ionizing
background from galaxies having other star-formation histories will be
explored.  However, such a general rise is easily reproduced when
``unfiltered'' spectra of stars are used in the CLOUDY modelling.
Adding the expected metagalactic QSO background reduces but does not
destroy this rise.  In the specific illustration of this in Figure 2 a
stellar contribution equivalent to about $\sim 30$ times the general
QSO flux at 1 Rydberg is indicated at the highest redshifts.  A
similar pattern of evolutionary behaviour is seen from a display of
N(Si II)/N(Si IV) vs N(C II)/N(C IV), which does not have the
complication of relative abundance differences.  Adding the same
stellar flux level at the low redshifts results in a very poor fit to
the data.

\section{Conclusions}

This all is indicating that at lower redshifts the ionization of the
metal systems is dominated by QSO light as modified by its passage
through the intergalactic medium, while at higher redshifts the
ionization becomes progressively dominated by the intense light of ``local''
stellar regions close enough to retain their intrinsic spectral shape (indeed increasing emphasis on local sources is expected at higher redshifts due to the increasing cosmic opacity).  
This has important bearing on the star formation history of the
Universe and can give evidence of galaxy evolution for a population of
structures not directly observable.  It is also suggestive of
an origin for the weak metal absorption systems of the Lyman forest
which is not related to Population III stars but more directly with
protogalactic or other star-forming structures.

\acknowledgements{We thank Bob Carswell for providing VPFIT, David
Valls-Gabaud and Roderick Johnstone for invaluable help with CLOUDY,
Francesco Haardt for providing many new computations of radiative
transfer models for the UVB, and Martin Haehnelt, David Valls-Gabaud
and Piero Madau for very helpful discussions.  }


\end{document}